\documentclass[sigconf]{acmart}
\usepackage{xcolor}
\newcommand{\bad}[1]{\textcolor{red}{\textbf{#1}}}
\usepackage{multirow}
\AtBeginDocument{%
  }

\setcopyright{acmlicensed}
\copyrightyear{2026}
\acmYear{2026}
\setcopyright{cc}
\setcctype{by}
\acmConference[AI Summit '26]{ACM AI Summit 2026}{August 31-September 02, 2026}{Atlanta, GA, USA}
\acmBooktitle{ACM AI Summit 2026 (AI Summit '26), August 31-September 02, 2026, Atlanta, GA, USA}
\acmDOI{10.1145/3806096.3844841}
\acmISBN{979-8-4007-2638-5/2026/08}

\begin{document}

\title{LadderTeam: Dual-Agent Laddering Elicitation Framework}


\author{Manjushree Aithal, PhD}
\affiliation{%
  \institution{University of Colorado Anschutz}
  \city{Aurora}
  \country{United States}}
\email{manjushree.aithal@cuanschutz.edu}

\author{Alexander Kotz}
\affiliation{%
  \institution{University of Colorado Anschutz}
  \city{Aurora}
  \country{United States}}
\email{alexander.kotz@cuanschutz.edu}

\author{James Mitchell, PhD}
\affiliation{%
  \institution{University of Colorado Anschutz}
  \city{Aurora}
  \country{United States}}
\email{james.2.mitchell@cuanschutz.edu}

\renewcommand{\shortauthors}{Aithal et al.}

\begin{abstract}

Eliciting detailed and actionable software requirements from end-users is a critical phase in the iterative development of a software product or application. To ensure the feedback collected is detailed and actionable, software teams can leverage the laddering interview technique. While effective for ensuring granular and actionable items from the software feedback, these interviews are subject to several limitations. They are traditionally a manual process associated with a time and financial burden, limiting scalability; interviewers must balance probing for depth while managing interviewee behavioral and cultural constraints. To address these limitations, we present \textbf{LadderTeam}, an open, reproducible framework that automates UX wireframe interviews using a dual-agent Large Language Model (LLM) architecture. An active interviewer agent executes one of three probing strategies (ACV, 5-Whys, and JTBD) to elicit actionable software requirements from usability feedback comments, while a concurrent background Judge agent evaluates probe-response pairs and triggers real-time guardrails to prevent topic drift. To rigorously evaluate LLM laddering without participant variance confounds, we introduce a controlled simulation methodology utilizing scripted ground-truth transcripts to isolate probe quality as the sole experimental variable. Across 216 interviews, \textbf{LadderTeam} achieved 99.1\% chain convergence and an 81.0\% ground-truth actionable response match (86.1\% reluctant personality, 75.9\% terse personality) with zero drift across all runs. \url{https://github.com/TheMitchellLab/Laddering-Module}
\end{abstract}



\begin{teaserfigure}
 \includegraphics[width=\textwidth]{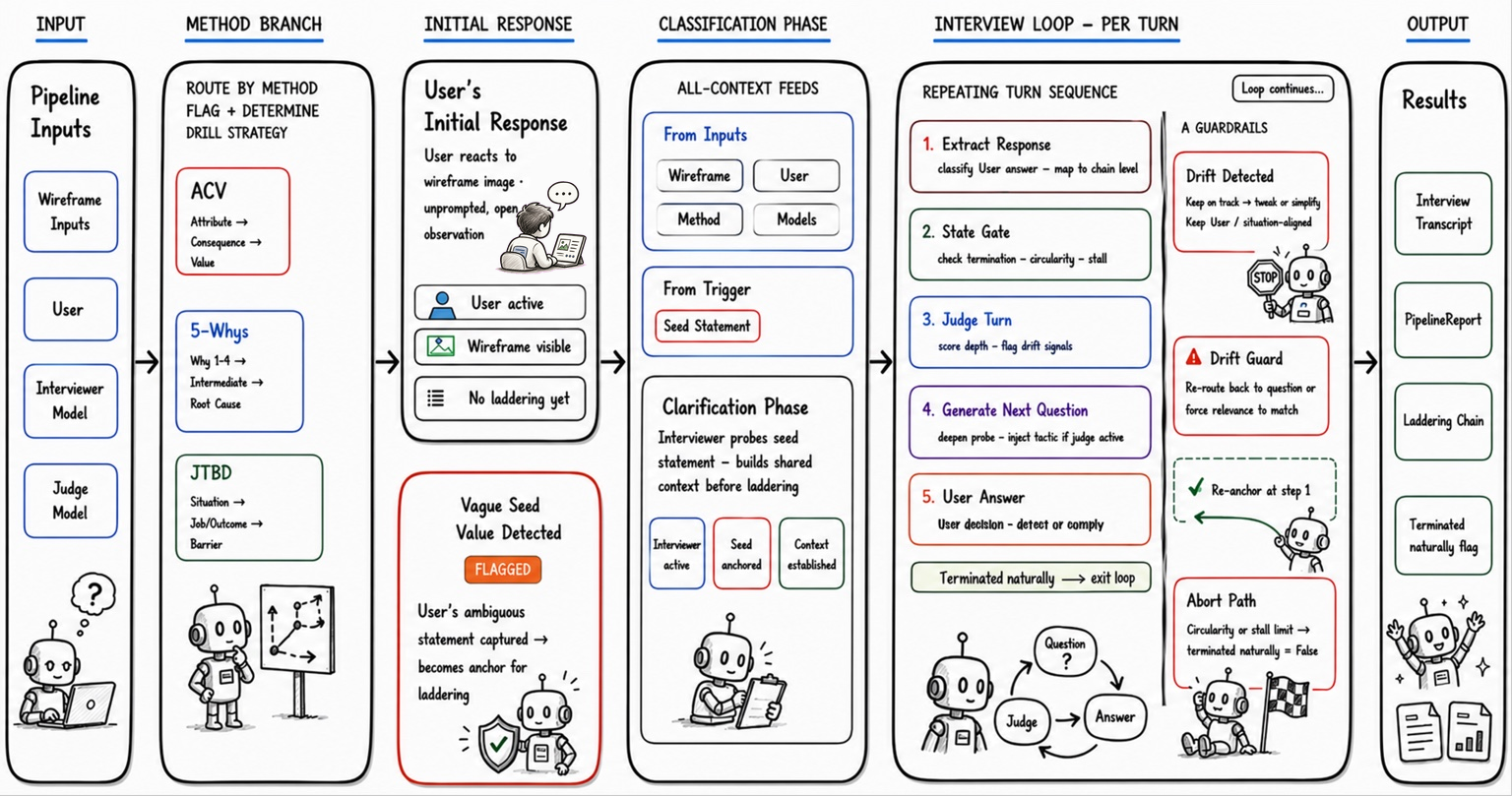}
  \caption{\textsc{LadderTeam} system overview. \textbf{Left:} wireframe + user + models+ method branch (ACV / 5-Whys / JTBD). \textbf{Center:} vague seed extraction
  from the prior response and multiple-choice questions (MCQ) clarification phase. \textbf{Right:} five-step
  per-turn loop with Drift Guard and Abort Path guardrails. Natural termination
  generates the interview transcript, chain, and \texttt{PipelineReport}. Note: Restyling of the image was performed using AI.}
  \Description{System overview of LadderTeam. Left panel: wireframe screens, user, and LLM models feeding into one of three method branches (ACV, 5-Whys, JTBD). Center panel: vague seed extraction from the user's prior response followed by an MCQ clarification phase. Right panel: five-step per-turn loop comprising Extract, State Gate, Judge, Generate Question, and User Response steps, with Drift Guard and Abort Path guardrails. Natural termination produces the interview transcript, laddering chain, and PipelineReport.}
  \label{fig:teaser}
\end{teaserfigure}

\maketitle

\section{Introduction}

Software usability feedback is critical to identify early bugs, assess overall usability, and collect ideas for new features~\cite{abedini_hybrid_2025, maalej_automated_2024}. While incredibly valuable, this feedback is often too vague for software developers, designers, or engineers to directly act on~\cite{folstad_users_2017, yusop_reporting_2017, galavi_online_2023}, often leaving responses such as, \textit{"This alert is useless"} or \textit{"The design is good"}. To collect usability feedback on software or applications, software teams leverage several common data collection techniques, including, but not limited to surveys or questionnaires~\cite{brooke_sus_1996}, focus groups (both in-person and online)~\cite{kontio_using_2004, galavi_online_2023}, and one-on-one interviews~\cite{rietz_ladderbot_2019, folstad_ladderchat_2025, maramba_methods_2019}. Surveys/questionnaires offer the advantage of scalable data collection, while focus groups offer the advantage of group ideation, but both lack a clear and effective method for clarifying vagueness in real-time. A common one-on-one interview technique, Laddering, addresses this limitation. Laddering is a structured interviewing technique that uses a series of directed probes to progressively surface the relationships between a user's stated problem, its functional implications, and the underlying goals or values driving them (e.g., a respondent states "The population health table on the dashboard is not surfacing the right data," which ladders up to a functional requirement of "Add height, weight, and BMI to the population health table" and an underlying value of "ensuring care management metrics are met for billing") — moving beyond what users volunteer upfront to uncover the richer structure of what they actually need~\cite{rugg_laddering_1995, reynolds_laddering_1988, corbridge_laddering_1994}. Laddering has been implemented as a knowledge elicitation interview technique across many domains including collecting medical student attitudes towards specific professional behaviors favored in medical doctors~\cite{miles_identifying_2010}, understanding effective teaching qualities of lectures in a university setting~\cite{voss_service_2007}, and collecting software requirements from users of an application (simulated through collecting common smartphone use cases)~\cite{rietz_ladderbot_2019}. More recently, Hanschmann et al.~\cite{folstad_ladderchat_2025} developed a conversational agent, LadderChat, to automate the Laddering interview process by tasking an LLM to act as the interviewer and collect information from participants on the topic of smartwatches. While preliminary, this work supports the use of LLMs to act as interviewers in Laddering interviews, which greatly reduces the manual burden of collecting software feedback and ensuring no vagueness exists in feedback~\cite{hamalainen_evaluating_2023, argyle_out_2023}. 

We present \textsc{LadderTeam}, an LLM-based scaffolding system for structured laddering interviews with users in software usability feedback contexts. Our study is scoped to this domain: participants view wireframe screens of a software product and respond to an LLM-moderated interview, with all methods, extractors, and Judge rubrics instantiated for software usability feedback. Our contributions are: 
(1)~a unified interview engine supporting ACV, 5-Whys, and JTBD for software usability
feedback, with a shared laddering seed-extraction and clarification phase and per-method state machine;
(2)~a reproducible five-step per-turn loop with explicit state gating, drift detection, and abort logic;
(3)~a background LLM Judge agent with Drift Guard and Abort Path guardrails that operate
without interrupting the conversational flow; and
(4)~preliminary evidence of high chain convergence rates and a working implementation of active Judge injection with Drift Guard and Abort Path guardrails across all three probing methods.

\section{Methods \& Framework}

LadderTeam employs a dual-agent architecture comprising two distinct LLM roles (Figure~\ref{fig:teaser}): an \emph{Interviewer} that actively conducts the user probing session and a background \emph{Judge} that silently evaluates every probe-response pair without interrupting the flow. To prevent evaluative bias from a fixed Judge, roles are reciprocally assigned: for cloud models, GPT-5.5 serves as Interviewer with Claude Sonnet~4.6 as Judge and vice-versa (local: Gemma4:12B/Qwen3.6:27B swapped accordingly).

\subsection{Pre-Interview Setup}


Consider a typical software usability evaluation where a participants review wireframes and share initial, unprompted impressions, which are often vague (e.g., ``Something feels off'') and lack the actionable specificity design teams require, thus triggering the LadderTeam interview protocol.

Once a vague initial response is received, the facilitator (human actor) identifies a \emph{laddering seed} from the initial response, which is a phrase subject to multiple interpretations. This seed, along with the initial response, wireframe image, and selected probing method, is passed to the Interviewer as the entry point for the session.

The Interviewer begins with a mandatory clarification round, presenting the user with a four-option multiple-choice question to firmly anchor the conversational topic before laddering commences. The user's selection is logged as ``confirmed-concern'' and persists across all subsequent turns to prevent topic drift. A DecayingMemoryBuffer ($\lambda = 0.85$, max 20 turns) maintains the session history, caching the prior response. 

\subsection{Probing Method}\label{sec:probing}
The system routes the session through one of three probing methodologies, each governed by a dedicated extractor, questioner, and state machine (Table~\ref{tab:methods}). While we evaluated these methods independently to isolate their performance characteristics, in practice, the facilitator can select the probing method that best aligns with their specific research objectives.

\begin{table}[t]
\caption{Summary of the three probing methodologies. Label Description in Section~\ref{sec:probing}}
\label{tab:methods}
\begin{tabular}{@{}p{1.5cm}p{1.5cm}p{1.8cm}p{1.7cm}@{}}
\toprule
Property               & ACV~\cite{gutman_means-end_1982}            & 5-Whys~\cite{ohno_toyota_2019}            & JTBD~\cite{ulwick_what_2005}              \\
\midrule
Labels      & A, C, V        & INT, ROOT         & S, J, O, B        \\
Chain order & A$\to$C$\to$V & causal depth   & S$\to$J$\to$O$\to$B \\
Termination & $\geq$1 each; C$\prec$V; d$\geq$3 & ROOT or depth & $\geq$1 each; S$\prec$J; d$\geq$3 \\
Cycling guard & C $\geq$2 & 3-tier circularity & O $\geq$3 \\
State class & \texttt{ChainState} & \texttt{FiveWhysState} & \texttt{JTBDState} \\
\bottomrule
\end{tabular}
\end{table}

\paragraph{ACV} is where the \texttt{ChainState} tracks observable UI elements [A] - the \textit{attribute}, functional or experiential UX impacts [C] - the \textit{consequence}, and actionable design requirements [V] - the \textit{value}~\cite{goodwin_designing_2011}. The value must be a concrete requirement (or set of requirements) that software teams can implement. The extractor is depth-aware at $C \geq 3$. Any required design directive is classified as~[V].

\paragraph{5-Whys} is where the \texttt{FiveWhysState} maps a causal chain utilizing a semantic question ban and three-tier circularity detection. Responses indicating a domain departure are automatically promoted to ``root-cause'' at a depth~$\geq$3.

\paragraph{JTBD (Jobs To Be Done)} is where the \texttt{JTBDState} isolates the situation or trigger [S], a solution-free functional objective [J], an Outcome-Driven Innovation outcome [O], and the underlying design friction [B]. The questioner strictly targets ``next-expected-level'', and the extractor remains locked to this target. The final artifact takes a standardized Job-Story form where it states ``When [S], I want to [J], so I can [O], but [B].'' 

\subsection{Per-Turn Loop} \label{sec:loop}
Regardless of the active method, every probe and response exchange executes a rigid 5-step sequence (Figure~\ref{fig:teaser}) as follows:

(1)~\textbf{Extract} where an LLM classifies the user's answer into the method's specific label set as described in Table~\ref{tab:methods}, incorporating depth and target awareness.

(2)~\textbf{State Gate} where authoritative and non-LLM programmatic logic evaluates state completion, circularity, and stall conditions.

(3)~\textbf{Judge} where the background LLM scores the probe-response pair. When in active mode, the Judge's suggested tactic passes as a mandatory override into Step 4, otherwise scores are logged silently.

(4)~\textbf{Generate Question} where the Interviewer LLM formulates the next probe. In active mode, the Judge's feedback influences probe generation for the next round.

(5)~\textbf{User Response} where the user's response is appended to the DecayingMemoryBuffer and fed back into Step 1.

This loop repeats until the state machine registers natural termination or an Abort Path is triggered.

\subsection{Judge Agent \& Guardrails}
While the Interviewer drives the session, the background Judge evaluates each exchange to prevent probing degradation such as generic or repetitive questions. Operating silently, the Judge supports shadow, active, and skip configurations. Each evaluation produces a \texttt{JudgeFeedback} record detailing ladder and deflection scores, unlock proximity, repetition flags, and suggested tactics.

To maintain session integrity, the system utilizes structural guardrails. The Drift Guard injects a re-anchoring instruction upon detecting divergence from the confirmed concern, and an escalation flag fires after three consecutive identical tactics. This occurs if a probe asks about team workflows when the confirmed concern was specifically a UI layout issue. Finally, the programmatic State Gate triggers an Abort Path to forcefully exit the loop when identifying circularity or stall limits. Following the session, the system generates a comprehensive report detailing drift patterns, missed deepening opportunities, and the overall laddering efficiency.

\begin{table*}[t]
\caption{Evaluation results across 216 runs (2 GT-scripts(P1=Reluctant, P2=Terse) $\times$ 4 models $\times$ 3 methods $\times$ 3 UI scenarios $\times$ 3 iterations) where values are aggregated across 9 iterations per cell. $\uparrow$=higher is better; $\downarrow$=lower is better. \textcolor{red}{\textbf{Red values}} indicate poor performance. $^\dagger$Judge-dependent metrics reflect the specific model applied and may vary accordingly. Drift rate 0.00 across all runs.}
\label{tab:results}
\scriptsize
\begin{tabular*}{\textwidth}{@{\extracolsep{\fill}}ll ccccccc ccccccc}
\toprule
& & \multicolumn{7}{c}{\textbf{P1 — Reluctant}} & \multicolumn{7}{c}{\textbf{P2 — Terse}} \\
\cmidrule(lr){3-9}\cmidrule(lr){10-16}
Method & Model
  & Conv.$\uparrow$ & GT$\uparrow$ & Turns$\downarrow$ & $\eta$$\uparrow$ & Defl.$^\dagger$$\downarrow$ & Rep.$^\dagger$$\downarrow$ & Esc.$^\dagger$$\downarrow$
  & Conv.$\uparrow$ & GT$\uparrow$ & Turns$\downarrow$ & $\eta$$\uparrow$ & Defl.$^\dagger$$\downarrow$ & Rep.$^\dagger$$\downarrow$ & Esc.$^\dagger$$\downarrow$ \\
\midrule
\multirow{4}{*}{ACV}
 & Gemma4:12B     & 9/9 & 9/9          & 6.33 & 0.64 & 2.12 & 0.03 & 0/3
              & 9/9 & 9/9          & 5.56 & 0.73 & 2.66 & 0.02 & 0/3 \\
 & Qwen3.6:27B    & 9/9 & \bad{7/9}  & 5.11 & \bad{0.80} & 2.78 & 0.00 & 1/3
              & 9/9 & 9/9          & 5.22 & 0.79 & 3.00 & 0.00 & 0/3 \\
 & GPT-5.5    & 9/9 & 9/9          & 5.67 & 0.73 & 1.62 & 0.28 & 0/3
              & 9/9 & 7/9          & 4.56 & 0.90 & 1.96 & 0.19 & 0/3 \\
 & Sonnet~4.6 & 9/9 & 9/9          & 5.45 & 0.76 & 2.77 & 0.05 & 0/3
              & 9/9 & \bad{5/9}  & 4.33 & \bad{0.89} & 2.75 & 0.14 & 0/3 \\
\midrule
\multirow{4}{*}{5-Whys}
 & Gemma4:12B     & 9/9        & 9/9        & 7.00 & 0.59 & 2.68 & 0.14 & 1/3
              & 9/9        & 8/9        & 4.78 & 0.87 & 2.81 & 0.00 & 0/3 \\
 & Qwen3.6:27B    & \bad{7/9} & 7/9        & 9.00 & 0.48 & 2.68 & 0.09 & 3/3
              & 9/9        & \bad{6/9} & 4.00 & \bad{1.00} & 3.00 & 0.00 & 0/3 \\
 & GPT-5.5    & 9/9        & \bad{6/9} & 7.11 & 0.60 & 1.95 & 0.28 & 0/3
              & 9/9        & \bad{6/9} & 4.00 & \bad{1.00} & 2.89 & 0.00 & 0/3 \\
 & Sonnet~4.6 & 9/9        & \bad{4/9} & 5.45 & 0.77 & 2.47 & 0.14 & 0/3
              & 9/9        & \bad{6/9} & 4.67 & \bad{0.87} & 2.90 & 0.04 & 0/3 \\
\midrule
\multirow{4}{*}{JTBD}
 & Gemma4:12B     & 9/9 & 9/9        & 6.78 & 0.59 & 2.02 & 0.11 & 0/3
              & 9/9 & \bad{5/9} & 7.67 & 0.53 & 1.96 & 0.02 & 0/3 \\
 & Qwen3.6:27B    & 9/9 & \bad{6/9} & 6.00 & 0.68 & 2.47 & 0.00 & 0/3
              & 9/9 & \bad{3/9} & 7.11 & 0.56 & 2.29 & 0.00 & 0/3 \\
 & GPT-5.5    & 9/9 & 9/9        & 6.22 & 0.65 & 2.11 & 0.00 & 0/3
              & 9/9 & 9/9        & 5.33 & 0.77 & 2.04 & 0.00 & 0/3 \\
 & Sonnet~4.6 & 9/9 & 9/9        & 6.56 & 0.62 & 2.30 & 0.38 & 0/3
              & 9/9 & 9/9        & 6.11 & 0.69 & 2.28 & 0.33 & 0/3 \\
\midrule
\multicolumn{2}{@{}l}{\emph{Overall}}
 & 106/108 & 93/108 & 6.39 & 0.66 & 2.33 & 0.12 & 5 grps
 & 108/108 & 82/108 & 5.28 & 0.80 & 2.54 & 0.06 & 0 grps \\
\bottomrule
\end{tabular*}
\end{table*}
\section{Results}

\subsection{Experimental Setup}
\subsubsection{General Setup}
Evaluation of the LadderTeam framework occurred in two stages. We first validated the end-to-end conversational flow, state transitions and natural termination across all three probing methods through an interview session involving one human and seven persona-based agents as interviewees. All sessions terminated naturally and produced actionable laddering chains, confirming the system operates as intended. Full transcripts from this stage will be released upon acceptance. Following this baseline establishment, the evaluation transitioned to scripted ground-truth transcripts to eliminate user response variance. By holding the user response as constant, the interviewer becomes the sole experimental variable. We executed 216 total interviews spanning four interviewer models, three probing methods, and three UI issue scenarios utilizing the P1= Reluctant \& P2= Terse (2 possible extreme cases) personality-type ground-truth transcripts~\cite{costa_neo_2014}. 

\subsubsection{Evaluation Metrics}
The following eight metrics are used for evaluation (*=Judge-dependent):

\begin{itemize}
  \item \textbf{Convergence~(Conv.)}: Binary value per run, 1 if the interview reached natural termination, 0 if the Abort Path fired; reported as $k/9$.

  \item \textbf{GT Response Match~(GT)}: Measures whether the interviewer elicited a response matching the ground-truth script, via normalized sequence ratio $r = 2M/T$ ($M$=matched chars, $T$=total chars); a turn passes if $r \geq 0.85$; reported as $k/9$.


  \item \textbf{Mean Turns~(Turns)}: Average turns to natural termination calculated as $\bar{t} = \frac{1}{3}\sum_{i=1}^{3} t_i$.

  \item \textbf{Laddering Efficiency~($\eta$)}: Measures ratio of minimum required depth to actual turns as $\eta = \min((d_{\min}+1)/\bar{t},\,1)$ where $d_{\min}=3$ for all three methods.

  \item \textbf{Deflection Score*~(Defl.)}: Mean of Judge assigned score ranging from 0 for direct answer to 3 for stall or loop.

  \item \textbf{Repetition Rate*~(Rep.)}: Proportion of reused probing tactics sourced from \texttt{repetition\_flag} in each turn's \texttt{JudgeFeedback} records.

  \item \textbf{Escalation*~(Esc.)}: A binary flag triggered when the Judge observed 3 consecutive identical tactics.

  \item \textbf{Drift Rate*}: Proportion of turns flagged by the Judge as departing from the confirmed concern.
\end{itemize}


\subsubsection{Curated Ground-truth Script}
For stage 2 of evaluation, the controlled simulation with pre-scripted transcripts dictating every user response turn-by-turn was utilized. These scripts encode two distinct deflection behaviors acting as strict probe quality gates. The STALL behavior activates when an interviewer issues a generic question without anchoring it in the user's words, causing the user to repeat a non-advancing response. The LOOP behavior activates when a probe fails to build upon a newly provided substantive answer, prompting the user to revert to earlier and more dismissive language. Effectively anchored probes advance the conversation while generic probes inevitably trigger a stall or loop. 

\subsection{Observations}
Table~\ref{tab:results} reports results across 216 runs. Interviewers reached a terminal chain state in 98.1\% for P1 scripted responses (106/108) \& 100\% for P2 responses (108/108). The only failure occurred in Qwen3.6:27B for 5-Whys, where successive why probes generated sufficient lexical overlap to prematurely trigger the state machine's circularity detector before root cause was confirmed. This is a probe quality failure rather than a structural depth ceiling. However, chain convergence does not guarantee reaching the actionable terminal value. The interviewer achieved 86.1\% and 75.9\% final value match against P1 and P2 respectively. The characteristic failure is high laddering efficiency $\eta=1.00$ paired with a low ground-truth match, indicating the interviewer accepted an intermediate chain node as the terminal instead of issuing the anchored probe required to elicit the ground-truth terminal response.

ACV proved the most reliable strategy against both scripts P1 \& P2, achieving 94.4\% against P1 and 83.3\% against P2. The 5-Whys \& JTBD methods tied at 72.2\% against P2. The 5-Whys ceiling reflects interviewers prematurely accepting brief causal answers as root-cause, while JTBD performance was limited by Qwen3.6:27B failing to pursue the B-level unlock. Cloud models (GPT-5.5 \& Sonnet~4.6) reached 9/9 ground-truth match ceilings on JTBD against P2, whereas local models diverged sharply, with Qwen3.6:27B dropping to 3/9. However, both local models recovered to 9/9 for ACV against P2. Escalation defined as 3 consecutive identical probe tactics was triggered exclusively on 5-Whys against P1 for Qwen3.6:27B and Gemma4:12B. However, against P2, interviews naturally concluded in 4 to 5 turns, leaving insufficient time for escalation.
 
 \section{Discussion}
The results demonstrate that controlled simulation effectively isolates failure modes obscured in uncontrolled environments, especially in qualitative research. The critical performance signal lies in the gap between chain convergence (99.1\%) and ground-truth match (81.0\%). This gap reveals that state machines often prematurely accept brief, non-specific answers, exposing a flaw invisible to convergence metrics. While zero drift confirms wireframe stimuli provide robust contextual anchoring, GPT-5.5 exposed a hidden issue: it achieved a perfect GT match through aggressive, repetitive questioning rather than careful probing, a behavioral flaw detectable only via deflection and repetition metrics. The selection of a probing method should align with a specific research objective. ACV is optimal for early-stage discovery because it extracts actionable design requirements suitable for immediate integration. Conversely, 5-Whys uncovers systematic root causes of known friction and JTBD validates mid-development functional goals. 

Future work of the framework extends across three primary directions. First, we will formally validate the Judge rubrics against human experts to ensure strict alignment with qualitative research standards. Second, we will aim to address the system's current sensitivity to vague seed input by exploring mechanisms that prevent weak inputs from generating shallow laddering chains. Finally, we plan to broaden the evaluation beyond single persona and wireframe to establish generalizability across diverse product domains, user demographics, and behavioral deflection styles.

\section{Conclusion}
We introduce \textbf{LadderTeam}, an automated, reproducible framework that scaffolds software usability laddering interviews by pairing an LLM interviewer with a concurrent background Judge. A controlled simulation evaluation across two personas yielded \textbf{99.1\% chain convergence and 81.0\% overall terminal response match}. ACV emerged as the most robust strategy across all simulations (94.4\% P1, 83.3\% P2), followed by JTBD (91.7\% P1, 72.2\% P2). However, a persistent premature termination failure mode demonstrated that convergence alone is an insufficient success metric. Ultimately, these findings position \textbf{LadderTeam} as an open, scalable LLM-assisted framework that resolves critical qualitative interview bottlenecks, empowering human experts to orchestrate deep qualitative UX research at a larger scale.

\section{Acknowledgement}
Research reported in this paper was supported by the National Library of Medicine (NLM) of the National Institutes of Health (NIH) under award number T15LM009451.
\bibliographystyle{ACM-Reference-Format}
\bibliography{main}



\end{document}